\documentclass[showpacs,aps,prl,twocolumn,amsmath,amssymb,epsfig,color]{
revtex4-1}

\usepackage{color}
\usepackage{graphicx}
\usepackage{epsfig}
\usepackage{dcolumn}
\usepackage{bm}
\newcommand     {\beq}[1]         { \begin{equation} #1 \end{equation} }

\begin{document}

\title{Rupture cascades in a discrete element model of a porous sedimentary
rock} 

\author{Ferenc Kun${}^{1}$\email{ferenc.kun@science.unideb.hu}, Imre
Varga${}^{1}$, 
Sabine Lennartz-Sassinek${}^{2,3}$, and Ian G.\ Main${}^{2}$} 
\affiliation{$^1$Department of Theoretical Physics, University of Debrecen,
P.O. Box 5, H-4010 Debrecen, Hungary}
\affiliation{$^2$School of Geosciences, University of Edinburgh,  EH9 3JW
Edinburgh, UK}
\affiliation{$^3$Institute for Geophysics and Meteorology, 
University of Cologne, Cologne, Germany}

\begin{abstract}
We investigate the scaling properties of the sources of crackling noise
in a fully-dynamic numerical model of sedimentary rocks
subject to uniaxial compression. The model is initiated by
filling a cylindrical container with randomly-sized spherical particles 
which are then connected by breakable beams. 
Loading at a constant strain rate the cohesive elements fail and 
the resulting stress transfer produces sudden bursts of correlated failures, 
directly analogous to the sources of acoustic emissions in real experiments. 
The source size, energy, and duration can all be quantified for an individual
event, and the population analyzed for their scaling properties, including the
distribution of waiting times between consecutive events.  Despite the
non-stationary loading, the results are all characterized by power law
distributions over a broad range of scales in agreement with experiments. 
As failure is approached temporal correlation of events emerge accompanied by
spatial 
clustering.
\end{abstract}
\maketitle

Understanding the processes that lead to catastrophic failure of porous granular
media is an important problem in a wide variety of applications, notably in
Earth science and engineering 
\cite{sammis_pag_1987,Biegel1989827,sammonds_role_1992,steacy_automaton_1991,
Heap201171,PhysRevLett.110.088702, 
vives2013_1,vives2013_2}.  
Such failure is often preceded by
detectable changes in mechanical properties (stress and strain), and in
geophysical signals (elastic wave velocity, electrical conductivity and acoustic
emissions) measured remotely at the sample boundary \cite{JGR:JGR11289}.  In
particular
acoustic emissions result from sources of internal damage due to sudden local
dislocations in the form of tensile or shear micro-cracks whose origin time,
location, orientation, duration, and magnitude can all be inferred from the radiated wave
train \cite{Graham2010161}. 
Typically only a very small proportion of the
micro-cracks revealed by destructive thin sectioning after the test results in
detectable acoustic emissions \cite{lockner_1993}.  As a consequence
experimental
data provide only a limited insight into the complexity of the microscopic
processes at work prior to failure, notably the probability distributions of the
relevant parameters, their scaling properties and their population dynamics. 

Theoretical approaches to the dynamics and statistics of rupture cascades 
have typically been based on stochastic fracture models comprising lattices of
springs \cite{PhysRevLett.108.225502}, 
beams \cite{carmona_fragmentation_2008,
PhysRevLett.104.095502}, fuses \cite{alava_role_2008,alava_statistical_2006}, or
fibers
\cite{nanjo_damage_2005,kun_universality_2008,pradhan_failure_2010}. 
However, such lattice models involve a strong
simplification of the material microstructure and the inhomogeneous stress
field.  For example macroscopic laws of damage for cohesive elements are often
implemented at the mesoscopic scale on a regular two dimensional grid, avoiding
the truly three dimensional microstructure of real porous media, and often using
power-law rheology as an input.   Here we adopt a discrete element modelling
(DEM) approach which relaxes all of these restrictions, and allows a realistic
investigation of the emergent properties of the dynamics, including the temporal
and spatial statistics of the resulting ‘crackling noise’. Starting from the
level of single particles of porous granular media, rupture cascades and scaling
laws both emerge spontaneously in the competition between realistic structural
disorder and the interactions and correlations that arise from
external dynamic loading and internal stress redistribution.  Our approach
quantitatively reproduces the observed scaling laws of crackling noise
remarkably well without tuning \cite{JGR:JGR11289,Graham2010161,Mair200025}, 
including those of parameters such
as burst energy and duration not available to lattice-based models.

\begin{figure}
\begin{center}
\epsfig{ bbllx=5,bblly=65,bburx=460,bbury=480,file=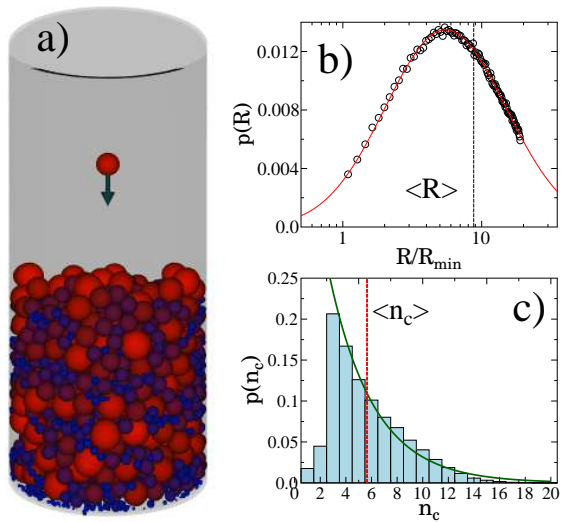,
width=7.7cm}
\end{center}
\caption{
(Color online) $(a)$ Preparation of the sample by sedimenting spherical
particles
with randomly distributed radius $R$.
The color code corresponds to the radius of the particles: the smallest
particles are dark blue while the biggest ones have light red color.
$(b)$ Comparison of the radius distribution $p(R)$ obtained numerically
(symbols) 
to the desired log-normal functional form (continuous line). $(c)$ Histogram 
$p(n_c)$ of the number of contacts $n_c$ in a sample of
$N=20000$ particles.
Exponential function (green line) was fitted for $n_c>3$.
}
\label{fig:sediment}
\end{figure}

In the model cylindrical samples are constructed  
by sedimenting spherical particles in a container.
Figure \ref{fig:sediment}$(a)$ illustrates that particles fall one-by-one 
on the top of the growing particle layer and dissipate their
kinetic energy by colliding with other particles and also with the 
container wall. 
The radius of particles $R$ was sampled from a log-normal distribution
$p(R) \sim \exp{\left[-(\ln R - \ln \overline{R})^2)/(2\sigma_R^2)\right]}$,
as shown in Fig.\ \ref{fig:sediment}$(b)$, which describes the statistics of
large 
particle sizes for various types of Earth materials 
(see e.g.\ the particle size distribution prior to faulting 
in Fig.\ 7 of Ref.\ \cite{Mair200025}). 
In order to avoid numerical problems of too wide size distributions, 
we set the range $R_{max}/R_{min}=20$ fixed and choose $\overline{R}=5R_{min}$
to have the maximum of $p(R)$ nearly in the middle of the $[\log R_{min}, \log
R_{max}]$
interval. The diameter $D_0$ and height $H_0$ of the cylinder were set to
$D_0=438.57R_{min}$ and $H_0=1008.71R_{min}$, which yields an aspect 
ratio $H_0/D_0\approx 2.3$ as in the experiments of Ref.\ \cite{Mair200025}.
With this geometrical setup the number of particles $N$  of the samples
fluctuates in a 
narrow interval around $N=20000$ with a total porosity $\Phi\approx 56\%$. 
Particles lying on the sample surface typically 
have only a few contacts $n_c=1,2,3$ to other ones, while bulk particles are
characterized
by higher contact numbers. The probability distribution (PDF) $p(n_c)$ of the
number 
of contacts $n_c$ proved to be an exponential for $n_c>3$, as
illustrated in Fig.\ \ref{fig:sediment}$(c)$.    
Both the exponential form of $p(n_c)$ and the value 
of the average number of contacts $\left<n_c\right> \approx 5.8$ of our packing 
are in a reasonable agreement with measurements on porous sandstones
\cite{Mair200025}.

To form a particulate solid in a DEM framework
\cite{potyondy_bonded-particle_2004,poschel_grandyn_2005,
carmona_fragmentation_2008,
PhysRevLett.104.095502}, 
cohesive interaction is provided by beams which connect 
the particles along the edges of a Delaunay tetrahedrization performed with the
initial 
position of particles. 
Conceptually the beam represents the effect of cementation and induration
between particles.
Beams can suffer elongation, compression, shear and 
torsion representing
the forces and torques which emerge between interacting particles  
\cite{poschel_grandyn_2005,carmona_fragmentation_2008,
PhysRevLett.104.095502}.
The time evolution of the system is followed by molecular dynamics simulations
solving the equation of motion of the particles. 
Beams break when overstressed, according to
\cite{carmona_fragmentation_2008,PhysRevLett.104.095502}
\beq{
\left(\frac{\varepsilon_{ij}}{\varepsilon_{th}}\right)^2 + 
\frac{max(|\Theta_i|, |\Theta_j|)}{\Theta_{th}} > 1,
\label{eq:breaking}
}
where $\varepsilon_{ij}$ denotes the axial strain, while $\Theta_i$ and
$\Theta_j$
are the generalized bending angles of the two ends of the beam connecting
particles
$i$ and $j$. The first and second terms of Eq.\ (\ref{eq:breaking}) represent
the contributions of stretching and bending,
respectively, 
where bending mainly arise due to the shear of the particle contacts 
\cite{daddetta_application_2002,carmona_fragmentation_2008,
PhysRevLett.104.095502}.
In the model there is only structural disorder present, i.e.\ the breaking
thresholds are set to constant values $\varepsilon_{th}=0.003$ and
$\Theta_{th}=2^{\rm o}$ for all the beams. 
Those particles which are not connected by beams (e.g.\ along cracks) interact 
via Hertz contacts \cite{poschel_grandyn_2005}.

\begin{figure}
\begin{center}
\epsfig{
bbllx=0,bblly=0,bburx=670,bbury=490,file=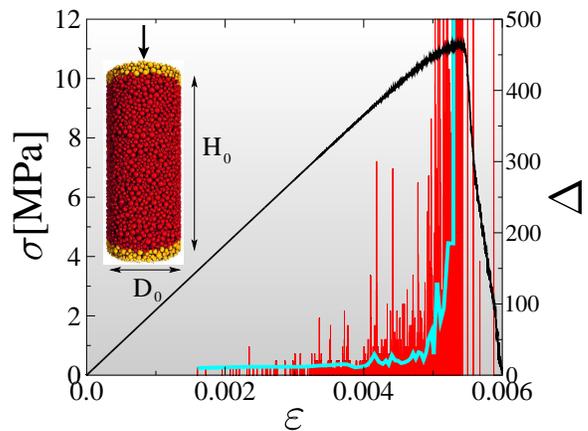,
width=8.2cm}
\end{center}
\caption{
(Color online) Constitutive behavior $\sigma(\varepsilon)$ of a single 
sample together with the series of burst size $\Delta$ (red bars) 
as a function of the strain of their appearance. The moving average of $\Delta$
(blue line) was calculated over 50 consecutive events.
The inset illustrates the loading condition. 
}
\label{fig:constit}
\end{figure}

To simulate uniaxial compression of sedimentary rocks in a strain controlled
way, 
two particle layers on the top and bottom of the cylindrical sample were clamped
such that the bottom layer was fixed while the one on the top was moving
downward at a constant speed (see the inset of Fig.\ \ref{fig:constit} for
illustration). 
The strain rate $\dot{\varepsilon}$ of loading was set as 
$\dot{\varepsilon}\Delta t = 1.8\cdot 10^{-7}$,
where $\Delta t$ is the time step used to integrate the equation of motion.
The constitutive curve $\sigma(\varepsilon)$ of the system is presented 
in Fig.\ \ref{fig:constit} where the measurement was stopped when the axial
stress
$\sigma$ dropped to zero. 
The system has a highly brittle response:
for small deformations linearly elastic behavior is obtained, 
stronger non-linearity of $\sigma(\varepsilon)$ is only observed 
in the vicinity of the maximum $\sigma_c$. 
Macroscopic failure is indicated by a sudden drop of the stress 
beyond the peak strain $\varepsilon_c$.

\begin{figure}
\begin{center}
\epsfig{
bbllx=20,bblly=5,bburx=740,bbury=635,file=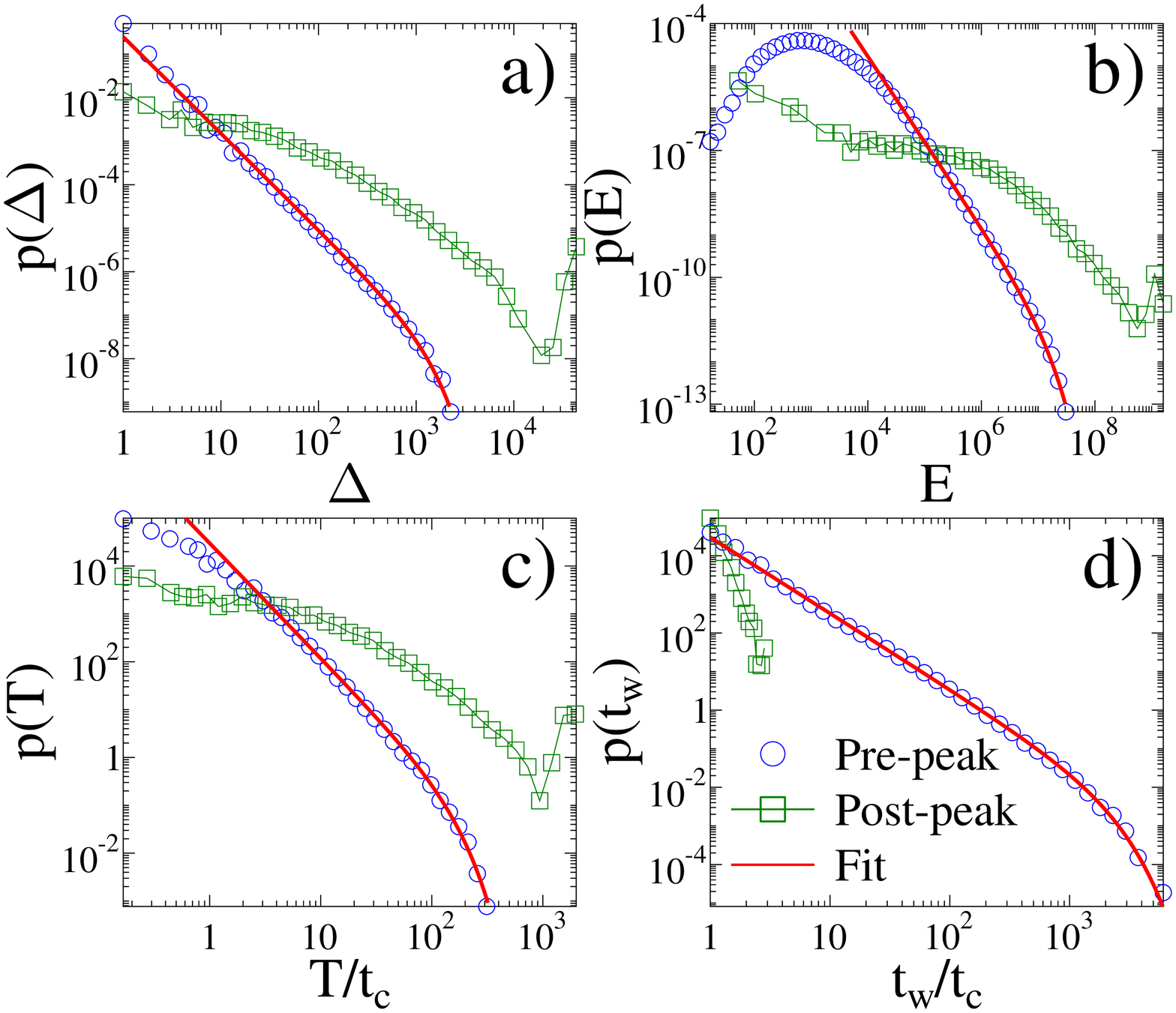,
width=8.1cm}
\end{center}
\caption{
(Color online) Probability distributions of the characteristic quantities
of bursts occurring before and after the peak of $\sigma(\varepsilon)$: 
distribution of burst $(a)$ size $p(\Delta)$, $(b)$ energy $p(E)$, 
$(c)$ duration $p(T)$, and $(d)$ waiting time $p(t_w)$. 
The red lines represent fits with the functional form of Eq.\
(\ref{eq:dist_delt}).
}
\label{fig:burstdistall}
\end{figure}
In the simulations the breaking criterion Eq.\ (\ref{eq:breaking}) 
is evaluated in each iteration step of the equation of motion
such that those beams which fulfill the condition are removed and  
their breaking time $t_i^b$ is recorded. 
During the loading process first the weakest beams break randomly  all
over the sample due to the quenched disorder starting at 
$\varepsilon \approx 0.0015$ in Fig.\ \ref{fig:constit}, i.e.\ relatively early 
in the loading history. 
Each breaking event is followed by the redistribution of stress which may induce
additional breakings and in turn can even trigger an entire avalanche of beam
breakings. 
If two consecutive beam breakings at times $t_i^b$ and $t_{i+1}^b$ occur 
within the correlation time $t_c$, i.e.\ $|t_{i+1}^b-t_i^b|<t_c$, they are
considered to
belong to the same burst. The value of $t_c$ was set to $t_c=25\Delta t$, which
is 
approximately the time needed for the elastic waves to pass the radius of the 
sample $D_0/2$. Similar criteria are also necessary to define real acoustic
emission
events in laboratory experiments
\cite{lockner_1993,sammonds_role_1992,Heap201171}. 
The breaking bursts of our DEM are
analogous to the acoustic emissions generated by the nucleation 
and propagation of cracks in laboratory experiments on geomaterials 
and in field observations on geological faults 
\cite{lockner_1993,sammonds_role_1992,Heap201171}.
We define the burst size $\Delta$ as the number of beams breaking in the
correlated sequence,
which is related to the rupture area created by the burst.
Figure \ref{fig:constit} shows that despite the smooth macroscopic response 
$\sigma(\varepsilon)$ of the system the size of bursts $\Delta$ exhibits
strong fluctuations while its average increases as the maximum of
$\sigma(\varepsilon)$ is approached. 
At the beginning of the breaking process only small bursts of a few breaking 
beams appear, however, as loading proceeds the triggering of longer 
avalanches becomes
more probable. Strong bursting activity with complex structure of the event
series
emerges after $\sigma$ exceeds approximately the two third of
the peak stress $\sigma_c$ in agreement with experiments \cite{Mair200025}. 
The total number of bursts we identify during the fracture
of a single sample is about 2000-2200.  

Since the dynamics of the breaking process changes at the peak load
we analyzed the statistical features of the time series of bursts separately for
events occurring
before $\varepsilon < \varepsilon_c$ and after $\varepsilon > \varepsilon_c$ the
peak 
of $\sigma(\varepsilon)$. 
Figure \ref{fig:burstdistall}$(a)$ shows that the PDF of 
burst sizes $p(\Delta)$ of pre-peak events has a power law functional form
followed
by a cutoff with stretched exponential shape
\beq{
p(\Delta) \sim \Delta^{-\tau}\exp{\left[-\left(\Delta/\Delta^{\ast}\right)^c
\right]}.
\label{eq:dist_delt}
}
A high quality fit was obtained with a rupture size exponent $\tau=2.22$, while
the cutoff
parameters are $c=1.5$ and $\Delta^{\ast}=1200$. 
At the peak of the constitutive curve $\sigma(\varepsilon)$ the dynamics of the
rupture process 
undergoes bifurcation, indicated by the different statistics of post-peak
events in Fig.\ \ref{fig:burstdistall}$(a)$. Although only a small fraction
$\sim 3\%$ of the bursts 
(about $12\%$ of broken beams) occurs along the softening branch of
$\sigma(\varepsilon)$,
large avalanches are more frequent in this regime. The small hump of the largest
events 
corresponds to the final multifragmentation of the sample. 
As the burst is formed, the elastic energy $E_j^b$ stored in beams is released, 
which can be directly compared to the energy
of acoustic signals in experiments. The overall duration $T$ of a burst 
is the difference of the time of the first and last beam breaking 
in the correlated sequence $T = t^b_{\Delta} - t^b_{1}$. Figures
\ref{fig:burstdistall}$(b)$
and $(c)$ show that in the pre-peak regime the PDF 
of burst energy $p(E)$ and duration $p(T)$ both have a power law decay with a
stretched
exponential cutoff similar to the behavior of the burst size Eq.\
(\ref{eq:dist_delt}).
Best fits were obtained with the power law exponents  $\alpha=2.02$, and
$\beta=2.4$, 
for the burst energy and duration, respectively, while the cutoff parameters are
$c_E=1.0$, $E^{\ast}=1.1 \times 10^{7}$ and $c_T=1.5$, and $T^{\ast}/t_c=170$. 
The corresponding distributions of
post-peak event source parameters (size, energy and duration) share a
similar qualitative shape, with a break of slope at low magnitude and a
bump at high values, the latter likely associated with the finite sample
size.

Bursts are separated by silent periods where no beam breaking occurs.
The duration $t_w$ of these inter-event periods 
encode interesting information about the temporal dynamics of fracture. 
The minimum value of $t_w$ is determined by the correlation
time $\min(t_w)\approx t_c$, hence, in Fig.\ 
\ref{fig:burstdistall}$(d)$ the PDF of waiting times $p(t_w)$ 
is presented as a function of the dimensionless ratio $t_w/t_c$ 
(the same is applied for $T$ in Fig.\ \ref{fig:burstdistall}$(c)$). 
Again the same functional form Eq.\ (\ref{eq:dist_delt}) of the distribution is
evidenced 
where best fit was obtained with the power law exponent
$z=2.0$, while the cutoff parameters are obtained as $c_w=1.2$ and 
$t_w^*/t_c=1600$. Note that separating post-peak events has only a minor 
effect on $p(t_w)$ since bursts in the post-peak regime rapidly occur with very
short
waiting times. We emphasize that the exponents $\tau, \alpha, \beta,$ and $z$ 
of the distributions are robust
with respect to the correlation time $t_c$ in the range $20\Delta t < t_c < 35
\Delta t$,
i.e.\ until $t_c$ falls close to the time the elastic wave takes to cross the
sample. Only 
the cutoffs of the distributions change. 
\begin{figure}
\begin{center}
\epsfig{
bbllx=1,bblly=170,bburx=755,bbury=775,
file=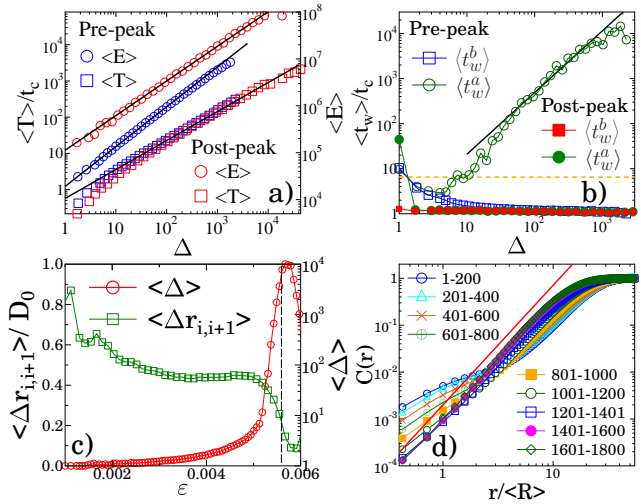,
width=8.5cm}
\end{center}
\caption{
(Color online) $(a)$ The average energy and duration of bursts as function of
their size. $(b)$ Average waiting times separately calculated 
before $\left<t_w^b\right>$ and after $\left<t_w^a\right>$ events as a function
of
$\Delta$. 
The straight lines in $(a)$ and $(b)$ are fits with Eqs.\
(\ref{eq:corre}, \ref{eq:corrtw}). $(c)$ Average distance between 
consecutive events and the average size of bursts as function of strain.
The vertical line indicates the peak position of $\sigma(\varepsilon)$.
$(d)$ The correlation integral $C(r)$ for windows of 200 consecutive events.
}
\label{fig:corrsizedurener}
\end{figure}

Characteristic quantities of single bursts $\Delta$, $E$, and $T$ are not
independent 
of each other: large bursts typically release a higher amount of energy and have
a 
longer duration. In order to quantify these correlations we determined the
average duration $\left<T\right>$ and energy $\left<E\right>$
of bursts as a function of their size $\Delta$ separately for pre- and post-peak
events. In Figure \ref{fig:corrsizedurener}$(a)$ a strong correlation is
observed 
with power law functional forms
\begin{eqnarray}
\label{eq:corre}
\left<E\right> \sim \Delta^{\nu_E}, \qquad  \mbox{and}  \qquad \left<T\right>
\sim \Delta^{\nu_T}.
\end{eqnarray}
The duration of bursts has the same behavior in both the pre- and post-peak
regimes with a unique exponent $\nu_T=0.8$. However, the energy of bursts 
of the same size proved to be higher for post- than for pre-peak events, 
since they are formed
by the breaking of stronger beams. The released energy is nearly proportional 
to the burst size with exponents $\nu_E=1.15$ and $\nu_E=1.0$ for 
$\varepsilon<\varepsilon_c$ and $\varepsilon>\varepsilon_c$, respectively.
These are lower than the scaling exponent of 1.5 commonly inferred from a
simple dislocation theory for the seismic source in interpreting
laboratory acoustic emission data \cite{hatton_1993}.
Equations (\ref{eq:corre}) yield relations between the pre-peak exponents 
$\alpha = (\tau + \nu_E -1)/\nu_E,$ and $\beta  = (\tau + \nu_T -1)/\nu_T$,
in good agreement with our numerically-determined exponents.

The stress
redistribution around cracks gives rise to correlations between bursts
which become more and more relevant as the system approaches
failure. 
To obtain information about how events affect 
the appearance of later bursts we determined the average value of waiting times 
as a function of the burst size $\Delta$ separately averaging $t_w$
that elapsed before $t_w^b$ and after $t_w^a$ the events.
Since along the softening branch of $\sigma(\varepsilon)$  
the specimen is collapsing with large bursts, in the post-peak regime
both $\left<t_w^a\right>$ and $\left<t_w^b\right>$ rapidly converge to
the vicinity of the most probable waiting time $t_c$ 
(see also Fig.\ \ref{fig:burstdistall}$(d)$) indicating
the absence of correlations. In the pre-peak regime $\left<t_w^b\right>$ has 
the same behavior though the convergence is slower.
The most remarkable result is that $\left<t_w^a\right>$ increases for large
event sizes
according to
\beq{
\left<t_w^a\right> \sim \Delta^{\nu_w},
\label{eq:corrtw}
}
with the exponent $\nu_w=1.37$. This correlation arises because a larger 
burst releases stress in a larger volume of the specimen so that it requires
a longer time to build up the stress again and to trigger the next burst. 
Our calculations revealed that the emergence of temporal correlations is also
accompanied by spatial 
clustering of events. Figure \ref{fig:corrsizedurener}$(c)$ presents the
average 
distance of consecutive bursts $\left<\Delta r_{i,i+1} \right>= 
\sqrt{\left<(\vec{r}_{i+1}-\vec{r}_i)^2 \right>}$ 
as a function of strain $\varepsilon$, where the position $\vec{r}_i$ of a
single burst
is identified by the center of mass of its broken beams. For a broad range of
$\varepsilon$
the ratio $\left<\Delta r_{i,i+1} \right>/D_0$ falls close to 0.5 which implies
that events randomly
scatter all over the sample. However, approaching the peak load $\sigma_c$ the
distance
rapidly decreases which clearly marks spatial clustering of events. In Fig.\
\ref{fig:corrsizedurener}$(c)$ 
the average size of bursts $\left<\Delta \right>$ increases with $\varepsilon$
and reaches a maximum
slightly beyond the peak of the consecutive curve. At the strain where spatial 
correlation sets on $\left<\Delta \right>$ switches to a faster growth. 
A more detailed measure of spatial correlation is provided by the correlation
integral $C(r)$
defined as $C(r)=N(<r)/N_p$, where $N(<r)$ denotes the number of pair of events
with a distance
smaller than $r$, and $N_p$ is the total number of pairs. To quantify how
correlations evolve,
we evaluated $C(r)$ for windows of 200 consecutive events. Fig.\
\ref{fig:corrsizedurener}$(d)$
shows that approaching failure the correlation integral saturates earlier and
for the 
last 4 windows it becomes a power law $C(r)\sim r^{D_2}$ with the
exponent $D_2=2.55$, which indicates strong spatial clustering of bursts.

In conclusion, we have successfully reconstructed a synthetic model of the
compressive failure of sedimentary rocks with realistic microstructure, breaking
dynamics and loading conditions relevant for catastrophic
failure in porous granular media. The statistical properties of
the local micro-crack events show qualitative agreement with those inferred from
acoustic emissions generated under compression in laboratory tests, notably the
power-law scaling of the PDFs of rupture area, duration and
energy and waiting time, and power-law scaling between rupture energy and
duration with respect to source size 
\cite{hatton_1993,vives2013_1,vives2013_2,ojala_2003,PhysRevLett.110.088702}.  
In recent laboratory experiments on porous rocks and on synthetic samples with
well
controlled porosity $\Phi$ power law distribution of the energy of acoustic
events was found 
with an exponent which increases with $\Phi$ from 1.6 to 2.0 
\cite{vives2013_1,vives2013_2}.
Our simulations have good qualitative
agreement with the time evolution of rupture \cite{vives2013_1,vives2013_2} 
and quantitative agreement with the energy exponent \cite{vives2013_2}. 
Our simulations also revealed microscopic details of the rupture process,
including the temporal evolution of spatial correlations in rupture location
that control the emergence of localized damage at a resolution not readily
accessible by experimental means, with potential implications for developing
predictive models for catastrophic failure in porous granular media.

\begin{acknowledgments}
We thank the projects TAMOP-4.2.2.A-11/1/KONV-2012-0036, 
TAMOP-4.2.2/B-10/1-2010-0024, OTKA K84157, and ERANET\_HU\_09-1-2011-0002.
This work was supported by the European Commissions by the
Complexity-NET pilot project LOCAT.
\end{acknowledgments}

\bibliography{paper_dem}

\begin{thebibliography}{10}%
\makeatletter
\providecommand \@ifxundefined [1]{%
 \ifx #1\undefined \expandafter \@firstoftwo
 \else \expandafter \@secondoftwo
\fi
}%
\providecommand \@ifnum [1]{%
 \ifnum #1\expandafter \@firstoftwo
 \else \expandafter \@secondoftwo
\fi
}%
\providecommand \enquote [1]{``#1''}%
\providecommand \bibnamefont  [1]{#1}%
\providecommand \bibfnamefont [1]{#1}%
\providecommand \citenamefont [1]{#1}%
\providecommand\href[0]{\@sanitize\@href}%
\providecommand\@href[1]{\endgroup\@@startlink{#1}\endgroup\@@href}%
\providecommand\@@href[1]{#1\@@endlink}%
\providecommand \@sanitize [0]{\begingroup\catcode`\&12\catcode`\#12\relax}%
\@ifxundefined \pdfoutput {\@firstoftwo}{%
 \@ifnum{\z@=\pdfoutput}{\@firstoftwo}{\@secondoftwo}%
}{%
 \providecommand\@@startlink[1]{\leavevmode\special{html:<a href="#1">}}%
 \providecommand\@@endlink[0]{\special{html:</a>}}%
}{%
 \providecommand\@@startlink[1]{%
  \leavevmode
  \pdfstartlink
   attr{/Border[0 0 1 ]/H/I/C[0 1 1]}%
   user{/Subtype/Link/A<</Type/Action/S/URI/URI(#1)>>}%
  \relax
 }%
 \providecommand\@@endlink[0]{\pdfendlink}%
}%
\providecommand \url  [0]{\begingroup\@sanitize \@url }%
\providecommand \@url [1]{\endgroup\@href {#1}{\urlprefix}}%
\providecommand \urlprefix [0]{URL }%
\providecommand \Eprint[0]{\href }%
\@ifxundefined \urlstyle {%
  \providecommand \doi [1]{doi:\discretionary{}{}{}#1}%
}{%
  \providecommand \doi [0]{doi:\discretionary{}{}{}\begingroup
  \urlstyle{rm}\Url }%
}%
\providecommand \doibase [0]{http://dx.doi.org/}%
\providecommand \Doi[1]{\href{\doibase#1}}%
\providecommand \bibAnnote [3]{%
  \BibitemShut{#1}%
  \begin{quotation}\noindent
    \textsc{Key:}\ #2\\\textsc{Annotation:}\ #3%
  \end{quotation}%
}%
\providecommand \bibAnnoteFile [2]{%
  \IfFileExists{#2}{\bibAnnote {#1} {#2} {\input{#2}}}{}%
}%
\providecommand \typeout [0]{\immediate \write \m@ne }%
\providecommand \selectlanguage [0]{\@gobble}%
\providecommand \bibinfo [0]{\@secondoftwo}%
\providecommand \bibfield [0]{\@secondoftwo}%
\providecommand \translation [1]{[#1]}%
\providecommand \BibitemOpen[0]{}%
\providecommand \bibitemStop [0]{}%
\providecommand \bibitemNoStop [0]{.\EOS\space}%
\providecommand \EOS [0]{\spacefactor3000\relax}%
\providecommand \BibitemShut [1]{\csname bibitem#1\endcsname}%
\bibitem{sammis_pag_1987}%
  \BibitemOpen
  \bibfield{author}{%
  \bibinfo {author} {\bibfnamefont{C.}~\bibnamefont{Sammis}}, \bibinfo {author}
  {\bibfnamefont{G.}~\bibnamefont{King}},\ and\ \bibinfo {author}
  {\bibfnamefont{R.}~\bibnamefont{Biegel}},\ }%
  \bibfield{journal}{%
  \bibinfo {journal} {Pure Appl. Geophys.}\ }%
  \textbf{\bibinfo {volume} {125}},\ \bibinfo {pages} {777} (\bibinfo {year}
  {1987})%
  \bibAnnoteFile{NoStop}{sammis_pag_1987}%
\bibitem{Biegel1989827}%
  \BibitemOpen
  \bibfield{author}{%
  \bibinfo {author} {\bibfnamefont{R.~L.}\ \bibnamefont{Biegel}}, \bibinfo
  {author} {\bibfnamefont{C.~G.}\ \bibnamefont{Sammis}},\ and\ \bibinfo
  {author} {\bibfnamefont{J.~H.}\ \bibnamefont{Dieterich}},\ }%
  \bibfield{journal}{%
  \bibinfo {journal} {J. Struct. Geology}\ }%
  \textbf{\bibinfo {volume} {11}},\ \bibinfo {pages} {827 } (\bibinfo {year}
  {1989})%
  \bibAnnoteFile{NoStop}{Biegel1989827}%
\bibitem{sammonds_role_1992}%
  \BibitemOpen
  \bibfield{author}{%
  \bibinfo {author} {\bibfnamefont{P.~R.}\ \bibnamefont{Sammonds}}, \bibinfo
  {author} {\bibfnamefont{P.~G.}\ \bibnamefont{Meredith}},\ and\ \bibinfo
  {author} {\bibfnamefont{I.~G.}\ \bibnamefont{Main}},\ }%
  \bibfield{journal}{%
  \bibinfo {journal} {Nature}\ }%
  \textbf{\bibinfo {volume} {359}},\ \bibinfo {pages} {228} (\bibinfo {year}
  {1992})%
  \bibAnnoteFile{NoStop}{sammonds_role_1992}%
\bibitem{steacy_automaton_1991}%
  \BibitemOpen
  \bibfield{author}{%
  \bibinfo {author} {\bibfnamefont{S.}~\bibnamefont{Steacy}}\ and\ \bibinfo
  {author} {\bibfnamefont{C.}~\bibnamefont{Sammis}},\ }%
  \bibfield{journal}{%
  \bibinfo {journal} {Nature}\ }%
  \textbf{\bibinfo {volume} {353}},\ \bibinfo {pages} {250–252} (\bibinfo
  {year} {1991})%
  \bibAnnoteFile{NoStop}{steacy_automaton_1991}%
\bibitem{Heap201171}%
  \BibitemOpen
  \bibfield{author}{%
  \bibinfo {author} {\bibfnamefont{M.}~\bibnamefont{Heap}}, \bibinfo {author}
  {\bibfnamefont{P.}~\bibnamefont{Baud}}, \bibinfo {author}
  {\bibfnamefont{P.}~\bibnamefont{Meredith}}, \bibinfo {author}
  {\bibfnamefont{S.}~\bibnamefont{Vinciguerra}}, \bibinfo {author}
  {\bibfnamefont{A.}~\bibnamefont{Bell}},\ and\ \bibinfo {author}
  {\bibfnamefont{I.}~\bibnamefont{Main}},\ }%
  \bibfield{journal}{%
  \bibinfo {journal} {Earth Planet. Sci. Lett.}\ }%
  \textbf{\bibinfo {volume} {307}},\ \bibinfo {pages} {71 } (\bibinfo {year}
  {2011})%
  \bibAnnoteFile{NoStop}{Heap201171}%
\bibitem{PhysRevLett.110.088702}%
  \BibitemOpen
  \bibfield{author}{%
  \bibinfo {author} {\bibfnamefont{J.}~\bibnamefont{Bar\'o}}, \bibinfo {author}
  {\bibfnamefont{A.}~\bibnamefont{Corral}}, \bibinfo {author}
  {\bibfnamefont{X.}~\bibnamefont{Illa}}, \bibinfo {author}
  {\bibfnamefont{A.}~\bibnamefont{Planes}}, \bibinfo {author}
  {\bibfnamefont{E.~K.~H.}\ \bibnamefont{Salje}}, \bibinfo {author}
  {\bibfnamefont{W.}~\bibnamefont{Schranz}}, \bibinfo {author}
  {\bibfnamefont{D.~E.}\ \bibnamefont{Soto-Parra}},\ and\ \bibinfo {author}
  {\bibfnamefont{E.}~\bibnamefont{Vives}},\ }%
  \bibfield{journal}{%
  \bibinfo {journal} {Phys. Rev. Lett.}\ }%
  \textbf{\bibinfo {volume} {110}},\ \bibinfo {pages} {088702} (\bibinfo {year}
  {2013})%
  \bibAnnoteFile{NoStop}{PhysRevLett.110.088702}%
\bibitem{vives2013_1}%
  \BibitemOpen
  \bibfield{author}{%
  \bibinfo {author} {\bibfnamefont{E.}~\bibnamefont{Salje}}, \bibinfo {author}
  {\bibfnamefont{G.}~\bibnamefont{Lampronti}}, \bibinfo {author}
  {\bibfnamefont{D.}~\bibnamefont{Soto-Parra}}, \bibinfo {author}
  {\bibfnamefont{J.}~\bibnamefont{Bar\'o}}, \bibinfo {author}
  {\bibfnamefont{A.}~\bibnamefont{Planes}},\ and\ \bibinfo {author}
  {\bibfnamefont{E.}~\bibnamefont{Vives}},\ }%
  \bibfield{journal}{%
  \bibinfo {journal} {American Mineralogist}\ }%
  \textbf{\bibinfo {volume} {98}},\ \bibinfo {pages} {609} (\bibinfo {year}
  {2013})%
  \bibAnnoteFile{NoStop}{vives2013_1}%
\bibitem{vives2013_2}%
  \BibitemOpen
  \bibfield{author}{%
  \bibinfo {author} {\bibfnamefont{P.}~\bibnamefont{Castillo-Villa}}, \bibinfo
  {author} {\bibfnamefont{J.}~\bibnamefont{Bar\'o}}, \bibinfo {author}
  {\bibfnamefont{A.}~\bibnamefont{Planes}}, \bibinfo {author}
  {\bibfnamefont{E.}~\bibnamefont{Salje}}, \bibinfo {author}
  {\bibfnamefont{P.}~\bibnamefont{Sellappan}}, \bibinfo {author}
  {\bibfnamefont{W.}~\bibnamefont{Kriven}},\ and\ \bibinfo {author}
  {\bibfnamefont{E.}~\bibnamefont{Vives}},\ }%
  \bibfield{journal}{%
  \bibinfo {journal} {arXiv:1305.3156}}%
   (\bibinfo {year} {2013})%
  \bibAnnoteFile{NoStop}{vives2013_2}%
\bibitem{JGR:JGR11289}%
  \BibitemOpen
  \bibfield{author}{%
  \bibinfo {author} {\bibfnamefont{C.~H.}\ \bibnamefont{Scholz}},\ }%
  \bibfield{journal}{%
  \bibinfo {journal} {J. Geophys. Res.}\ }%
  \textbf{\bibinfo {volume} {73}},\ \bibinfo {pages} {1417} (\bibinfo {year}
  {1968})%
  \bibAnnoteFile{NoStop}{JGR:JGR11289}%
\bibitem{Graham2010161}%
  \BibitemOpen
  \bibfield{author}{%
  \bibinfo {author} {\bibfnamefont{C.~C.}\ \bibnamefont{Graham}}, \bibinfo
  {author} {\bibfnamefont{S.}~\bibnamefont{Stanchits}}, \bibinfo {author}
  {\bibfnamefont{I.~G.}\ \bibnamefont{Main}},\ and\ \bibinfo {author}
  {\bibfnamefont{G.}~\bibnamefont{Dresen}},\ }%
  \bibfield{journal}{%
  \bibinfo {journal} {Int. J. Rock Mech. Min. Sci.}\ }%
  \textbf{\bibinfo {volume} {47}},\ \bibinfo {pages} {161 } (\bibinfo {year}
  {2010})%
  \bibAnnoteFile{NoStop}{Graham2010161}%
\bibitem{lockner_1993}%
  \BibitemOpen
  \bibfield{author}{%
  \bibinfo {author} {\bibfnamefont{D.}~\bibnamefont{Lockner}},\ }%
  \bibfield{journal}{%
  \bibinfo {journal} {Int. J. Rock Mech. Min. Sci. \& Geomech. Abstr.}\ }%
  \textbf{\bibinfo {volume} {30}},\ \bibinfo {pages} {883} (\bibinfo {year}
  {1993})%
  \bibAnnoteFile{NoStop}{lockner_1993}%
\bibitem{PhysRevLett.108.225502}%
  \BibitemOpen
  \bibfield{author}{%
  \bibinfo {author} {\bibfnamefont{L.}~\bibnamefont{Girard}}, \bibinfo {author}
  {\bibfnamefont{J.}~\bibnamefont{Weiss}},\ and\ \bibinfo {author}
  {\bibfnamefont{D.}~\bibnamefont{Amitrano}},\ }%
  \bibfield{journal}{%
  \bibinfo {journal} {Phys. Rev. Lett.}\ }%
  \textbf{\bibinfo {volume} {108}},\ \bibinfo {pages} {225502} (\bibinfo {year}
  {2012})%
  \bibAnnoteFile{NoStop}{PhysRevLett.108.225502}%
\bibitem{carmona_fragmentation_2008}%
  \BibitemOpen
  \bibfield{author}{%
  \bibinfo {author} {\bibfnamefont{H.~A.}\ \bibnamefont{Carmona}}, \bibinfo
  {author} {\bibfnamefont{F.~K.}\ \bibnamefont{Wittel}}, \bibinfo {author}
  {\bibfnamefont{F.}~\bibnamefont{Kun}},\ and\ \bibinfo {author}
  {\bibfnamefont{H.~J.}\ \bibnamefont{Herrmann}},\ }%
  \bibfield{journal}{%
  \bibinfo {journal} {Phys. Rev. E}\ }%
  \textbf{\bibinfo {volume} {77}},\ \bibinfo {pages} {051302} (\bibinfo {year}
  {2008})%
  \bibAnnoteFile{NoStop}{carmona_fragmentation_2008}%
\bibitem{PhysRevLett.104.095502}%
  \BibitemOpen
  \bibfield{author}{%
  \bibinfo {author} {\bibfnamefont{G.}~\bibnamefont{Tim\'ar}}, \bibinfo
  {author} {\bibfnamefont{J.}~\bibnamefont{Bl\"omer}}, \bibinfo {author}
  {\bibfnamefont{F.}~\bibnamefont{Kun}},\ and\ \bibinfo {author}
  {\bibfnamefont{H.~J.}\ \bibnamefont{Herrmann}},\ }%
  \bibfield{journal}{%
  \bibinfo {journal} {Phys. Rev. Lett.}\ }%
  \textbf{\bibinfo {volume} {104}},\ \bibinfo {pages} {095502} (\bibinfo {year}
  {2010})%
  \bibAnnoteFile{NoStop}{PhysRevLett.104.095502}%
\bibitem{alava_role_2008}%
  \BibitemOpen
  \bibfield{author}{%
  \bibinfo {author} {\bibfnamefont{M.~J.}\ \bibnamefont{Alava}}, \bibinfo
  {author} {\bibfnamefont{P.~K. V.~V.}\ \bibnamefont{Nukala}},\ and\ \bibinfo
  {author} {\bibfnamefont{S.}~\bibnamefont{Zapperi}},\ }%
  \bibfield{journal}{%
  \bibinfo {journal} {Phys. Rev. Lett.}\ }%
  \textbf{\bibinfo {volume} {100}},\ \bibinfo {pages} {055502} (\bibinfo {year}
  {2008})%
  \bibAnnoteFile{NoStop}{alava_role_2008}%
\bibitem{alava_statistical_2006}%
  \BibitemOpen
  \bibfield{author}{%
  \bibinfo {author} {\bibfnamefont{M.}~\bibnamefont{Alava}}, \bibinfo {author}
  {\bibfnamefont{P.~K.}\ \bibnamefont{Nukala}},\ and\ \bibinfo {author}
  {\bibfnamefont{S.}~\bibnamefont{Zapperi}},\ }%
  \bibfield{journal}{%
  \bibinfo {journal} {Adv. in Phys.}\ }%
  \textbf{\bibinfo {volume} {55}},\ \bibinfo {pages} {349–476} (\bibinfo
  {year} {2006})%
  \bibAnnoteFile{NoStop}{alava_statistical_2006}%
\bibitem{nanjo_damage_2005}%
  \BibitemOpen
  \bibfield{author}{%
  \bibinfo {author} {\bibfnamefont{K.~Z.}\ \bibnamefont{Nanjo}}\ and\ \bibinfo
  {author} {\bibfnamefont{D.~L.}\ \bibnamefont{Turcotte}},\ }%
  \bibfield{journal}{%
  \bibinfo {journal} {Geophys. J. Int.}\ }%
  \textbf{\bibinfo {volume} {162}},\ \bibinfo {pages} {859} (\bibinfo {year}
  {2005})%
  \bibAnnoteFile{NoStop}{nanjo_damage_2005}%
\bibitem{kun_universality_2008}%
  \BibitemOpen
  \bibfield{author}{%
  \bibinfo {author} {\bibfnamefont{F.}~\bibnamefont{Kun}}, \bibinfo {author}
  {\bibfnamefont{H.~A.}\ \bibnamefont{Carmona}}, \bibinfo {author}
  {\bibfnamefont{J.~S.}\ \bibnamefont{Andrade}},\ and\ \bibinfo {author}
  {\bibfnamefont{H.~J.}\ \bibnamefont{Herrmann}},\ }%
  \bibfield{journal}{%
  \bibinfo {journal} {Phys. Rev. Lett.}\ }%
  \textbf{\bibinfo {volume} {100}},\ \bibinfo {pages} {094301} (\bibinfo {year}
  {2008})%
  \bibAnnoteFile{NoStop}{kun_universality_2008}%
\bibitem{pradhan_failure_2010}%
  \BibitemOpen
  \bibfield{author}{%
  \bibinfo {author} {\bibfnamefont{S.}~\bibnamefont{Pradhan}}, \bibinfo
  {author} {\bibfnamefont{A.}~\bibnamefont{Hansen}},\ and\ \bibinfo {author}
  {\bibfnamefont{B.~K.}\ \bibnamefont{Chakrabarti}},\ }%
  \bibfield{journal}{%
  \bibinfo {journal} {Rev. Mod. Phys.}\ }%
  \textbf{\bibinfo {volume} {82}},\ \bibinfo {pages} {499} (\bibinfo {year}
  {2010})%
  \bibAnnoteFile{NoStop}{pradhan_failure_2010}%
\bibitem{Mair200025}%
  \BibitemOpen
  \bibfield{author}{%
  \bibinfo {author} {\bibfnamefont{K.}~\bibnamefont{Mair}}, \bibinfo {author}
  {\bibfnamefont{I.}~\bibnamefont{Main}},\ and\ \bibinfo {author}
  {\bibfnamefont{S.}~\bibnamefont{Elphick}},\ }%
  \bibfield{journal}{%
  \bibinfo {journal} {J. Struct. Geol.}\ }%
  \textbf{\bibinfo {volume} {22}},\ \bibinfo {pages} {25 } (\bibinfo {year}
  {2000})%
  \bibAnnoteFile{NoStop}{Mair200025}%
\bibitem{potyondy_bonded-particle_2004}%
  \BibitemOpen
  \bibfield{author}{%
  \bibinfo {author} {\bibfnamefont{D.}~\bibnamefont{Potyondy}}\ and\ \bibinfo
  {author} {\bibfnamefont{P.}~\bibnamefont{Cundall}},\ }%
  \bibfield{journal}{%
  \bibinfo {journal} {Int. J. Rock Mech. Min. Sci.}\ }%
  \textbf{\bibinfo {volume} {41}},\ \bibinfo {pages} {1329} (\bibinfo {year}
  {2004})%
  \bibAnnoteFile{NoStop}{potyondy_bonded-particle_2004}%
\bibitem{poschel_grandyn_2005}%
  \BibitemOpen
  \bibfield{author}{%
  \bibinfo {author} {\bibfnamefont{T.}~\bibnamefont{P\"{o}schel}}\ and\
  \bibinfo {author} {\bibfnamefont{T.}~\bibnamefont{Schwager}},\ }%
  \emph{\bibinfo {title} {{Computational Granular Dynamics}}}\ (\bibinfo
  {publisher} {Springer},\ \bibinfo {address} {Berlin},\ \bibinfo {year}
  {2005})%
  \bibAnnoteFile{NoStop}{poschel_grandyn_2005}%
\bibitem{daddetta_application_2002}%
  \BibitemOpen
  \bibfield{author}{%
  \bibinfo {author} {\bibfnamefont{G.~A.}\ \bibnamefont{{D'Addetta}}}, \bibinfo
  {author} {\bibfnamefont{F.}~\bibnamefont{Kun}},\ and\ \bibinfo {author}
  {\bibfnamefont{E.}~\bibnamefont{Ramm}},\ }%
  \bibfield{journal}{%
  \bibinfo {journal} {Granular Matter}\ }%
  \textbf{\bibinfo {volume} {4}},\ \bibinfo {pages} {77} (\bibinfo {year}
  {2002})%
  \bibAnnoteFile{NoStop}{daddetta_application_2002}%
\bibitem{hatton_1993}%
  \BibitemOpen
  \bibfield{author}{%
  \bibinfo {author} {\bibfnamefont{C.~G.}\ \bibnamefont{Hatton}}, \bibinfo
  {author} {\bibfnamefont{I.~G.}\ \bibnamefont{Main}},\ and\ \bibinfo {author}
  {\bibfnamefont{P.~G.}\ \bibnamefont{Meredith}},\ }%
  \bibfield{journal}{%
  \bibinfo {journal} {J. Struct. Geol.}\ }%
  \textbf{\bibinfo {volume} {15}},\ \bibinfo {pages} {1485} (\bibinfo {year}
  {1993})%
  \bibAnnoteFile{NoStop}{hatton_1993}%
\bibitem{ojala_2003}%
  \BibitemOpen
  \bibfield{author}{%
  \bibinfo {author} {\bibfnamefont{I.}~\bibnamefont{Ojala}}, \bibinfo {author}
  {\bibfnamefont{B.~T.}\ \bibnamefont{Ngwenya}}, \bibinfo {author}
  {\bibfnamefont{I.~G.}\ \bibnamefont{Main}},\ and\ \bibinfo {author}
  {\bibfnamefont{S.~C.}\ \bibnamefont{Elphick}},\ }%
  \bibfield{journal}{%
  \bibinfo {journal} {J. Geophys. Res.}\ }%
  \textbf{\bibinfo {volume} {108 (B5)}},\ \bibinfo {pages} {2268} (\bibinfo
  {year} {2003})%
  \bibAnnoteFile{NoStop}{ojala_2003}%
\end{thebibliography}%

\end{document}